\documentclass[preprint,aps,showkeys,amsmath,amssymb,nopreprintnumbers,nofootinbib]{revtex4}
\usepackage{graphicx}
\usepackage{dcolumn}
\usepackage{bm}

\def\bge{\begin{equation}}
\def\ene{\end{equation}}
\def\bg{\begin{eqnarray}}
\def\en{\end{eqnarray}}

\begin{document}


\title{Effects of Fock term, tensor coupling and baryon structure variation on a neutron star}%

\author{Tsuyoshi Miyatsu, Tetsuya Katayama and Koichi Saito}
\email{koichi.saito@rs.tus.ac.jp}
\affiliation{%
Department of Physics, Faculty of Science and Technology,\\
Tokyo University of Science, Noda 278-8510, Japan 
}%


\date{\today}

\begin{abstract}
The equation of state for neutron matter is calculated within relativistic Hartree-Fock approximation. 
The tensor couplings of vector mesons to baryons are included, and the change of baryon internal structure in matter is also considered using the quark-meson coupling model. 
We obtain the maximum neutron-star mass of $\sim 2.0 M_\odot$, which is consistent with the recently observed, precise mass, $1.97\pm0.04 M_{\odot}$. 
The Fock contribution is very important and, in particular, the inclusion of tensor coupling is vital to obtain such large mass. 
The baryon structure variation in matter also enhances the mass of a neutron star. 
\end{abstract}

\maketitle

Because the internal structure and the mass of a neutron star significantly depend on the equation of state (EOS) for neutron matter, the observation of pulsars provides important information to understand the properties of dense matter~\cite{Glendenning:1997wn}. 
At sufficiently high density, nuclear or neutron matter would be composed of not only nucleons ($N$) and leptons ($\ell$) but also hyperons ($Y$)~\cite{Glendenning:1997wn,Schaffner:1995th,Pal:1999sq,Shen:2002qg,Ishizuka:2008gr} and, possibly, meson condensates~\cite{Schaffner:1995th,Meson-condensation}. 
Since it is important to calculate the EOS relativistically, many studies on neutron matter have been performed within relativistic Hartree approximation.  

Recently, the precise observation finds the pulsar with the mass of $1.97\pm0.04 M_{\odot}$ (solar mass)~\cite{Demorest:2010bx}.  
It is, however, difficult to explain the observed mass by the EOS calculated in relativistic Hartree approximation, because the inclusion of hyperons makes the EOS soft and thus reduces the maximum mass of a neutron star~\cite{Pal:1999sq,Shen:2002qg,Ishizuka:2008gr}.  
Thus, it seems very urgent to consider how this discrepancy can be reconciled. It may be necessary to first study the effects of the Fock term, the tensor couplings of vector mesons and the baryon structure change in matter, and see how  those effects contribute to the EOS or the maximum mass of a neutron star. 

In fact, the relativistic Hartree-Fock calculation of the EOS for neutron matter has already been performed 
in Quantum Hadrodynamics (QHD)~\cite{RHF-Huber} or 
in the quark-meson coupling (QMC) model~\cite{RikovskaStone:2006ta}, and it is 
pointed out that the Fock term is significant to enhance the maximum mass. 
However, in Ref.~\cite{RHF-Huber}, the effect of the baryon structure change in matter is ignored.  
In Ref.~\cite{RikovskaStone:2006ta}, although the maximum mass of 
about $2 M_{\odot}$ is obtained, their calculation misses the tensor coupling and 
the space component of baryon self-energy.  

In this paper, we study the properties of nuclear and neutron matter 
within relativistic Hartree-Fock approximation~\cite{Serot:1984ey,Bouyssy:1987sh}, where 
the pion exchange and the tensor couplings of vector mesons to 
baryons are included in the Fock term.  
Furthermore, using the QMC model~\cite{QMC}, 
we consider the variation of the quark substructure of baryon in matter. 
In the past few decades, the QMC model has been extensively developed and applied to various nuclear phenomena 
with tremendous success~\cite{QMC}.  
We also apply the chiral QMC (CQMC) model~\cite{QMC3,Saito:2010zw}, 
which is recently extended to include the quark-quark hyperfine interaction 
due to the gluon~\cite{Guichon:2008zz} and pion exchanges based on chiral symmetry, to the EOS for neutron matter. 

The Lagrangian density for dense matter is given by
\bge
	\mathcal{L} = \mathcal{L}_{B} + \mathcal{L}_{\ell} + \mathcal{L}_{M} + \mathcal{L}_{int} ,
	\label{eq:Lagrangian1}
\ene
where
\bge
	\mathcal{L}_{B} = \sum_{B}\bar{\psi}_{B}(i\gamma_{\mu}\partial^{\mu}-M_{B})\psi_{B}, \ \ \ \ \ 
	\mathcal{L}_{\ell} = \sum_{\ell}\bar{\psi}_{\ell}(i\gamma_{\mu}\partial^{\mu}-m_{\ell})\psi_{\ell},
	\label{eq:Lagrangian-baryon-lepton}
\ene
with $\psi_{B (\ell)}$ the baryon (lepton) field and $M_{B} (m_{\ell})$ the free baryon (lepton) mass. 
The sum $B$ runs over the octet baryons, $p$, $n$, $\Lambda$, $\Sigma^{+0-}$ and $\Xi^{0-}$, 
and the sum $\ell$ is for the leptons, $e^{-}$ and $\mu^{-}$.  For the baryon masses, we take $M_{N}=939$ MeV, 
$M_{\Lambda}=1116$ MeV, $M_{\Sigma}=1193$ MeV and $M_{\Xi}=1313$ MeV, respectively.

The meson term reads
\bg
	\mathcal{L}_{M} &=& \frac{1}{2}\left(\partial_{\mu}\sigma\partial^{\mu}\sigma-m_{\sigma}^{2}\sigma^{2}\right)
	+ \frac{1}{2}m_{\omega}^{2}\omega_{\mu}\omega^{\mu} - \frac{1}{4}W_{\mu\nu}W^{\mu\nu} \nonumber \\
	&+& \frac{1}{2}m_{\rho}^{2}\vec{\rho}_{\mu}\cdot\vec{\rho}^{\,\mu} - \frac{1}{4}\vec{R}_{\mu\nu}\cdot\vec{R}^{\mu\nu}
	+ \frac{1}{2}\left(\partial_{\mu}\vec{\pi}\cdot\partial^{\mu}\vec{\pi}-m_{\pi}^{2}\vec{\pi}^{2}\right),
	\label{eq:Lagrangian-meson}
\en
with
\bge
	W_{\mu\nu} = \partial_{\mu}\omega_{\nu} - \partial_{\nu}\omega_{\mu}, \ \ \ \ \ 
	\vec{R}_{\mu\nu} = \partial_{\mu}\vec{\rho}_{\nu} - \partial_{\nu}\vec{\rho}_{\mu},
	\label{eq:covariant-derivative}
\ene
where we consider the isoscalar ($\sigma$ and $\omega$) mesons and 
the isovector (${\vec \pi}$ and ${\vec \rho\,}$) mesons, and 
the meson masses are respectively chosen as 
$m_{\sigma}=550$ MeV, $m_{\omega}=783$ MeV, $m_{\pi}=138$ MeV and $m_{\rho}=770$ MeV.  

The interaction Lagrangian is given by
\bg
	\mathcal{L}_{int} &=& \sum_{B}\bar{\psi}_{B}\left[g_{\sigma B}(\sigma)\sigma
	-g_{\omega B}\gamma_{\mu}\omega^{\mu} 
	+ \frac{f_{\omega B}}{2\mathcal{M}}\sigma_{\mu\nu}\partial^{\nu}\omega^{\mu}\right. \nonumber \\ 
	&-& \left. g_{\rho B}\gamma_{\mu}\vec{\rho}^{\,\mu}\cdot\vec{I}_B
	+ \frac{f_{\rho B}}{2\mathcal{M}}\sigma_{\mu\nu}\partial^{\nu}\vec{\rho}^{\,\mu}\cdot\vec{I}_B
	- \frac{f_{\pi B}}{m_{\pi}}\gamma_{5}\gamma_{\mu}\partial^{\mu}\vec{\pi}\cdot\vec{I}_B \right]\psi_{B},
	\label{eq:Lagrangian-interaction}
\en
where $\vec{I}_B$ is the isospin matrix for baryon $B$ (we set $\vec{I}_B=0$ when $B$ is iso-singlet) 
and the common, scale mass $\mathcal{M}$ is taken to be the free proton mass~\cite{Rijken:2010zzb}. 
The $\sigma$-, $\omega$-, $\rho$-, $\pi$-$B$ coupling constants are respectively denoted by 
$g_{\sigma B}(\sigma)$, $g_{\omega B}$, $g_{\rho B}$ and $f_{\pi B}$, 
while $f_{\omega B}$ and $f_{\rho B}$ are the isoscalar- and isovector-tensor coupling constants. 

In QHD, $g_{\sigma B}$ is constant at any density because the baryons are assumed to be structureless.  
In contrast, in the QMC and CQMC models, 
the $g_{\sigma B}(\sigma)$ has the $\sigma$-field dependence, 
which reflects the baryon structure variation in matter~\cite{QMC}. 
In relativistic Hartree-Fock approximation, 
the $\sigma$-field is replaced by the constant mean-field value, $\bar{\sigma}$. 
Furthermore, for simplicity, we here use the parameterization of 
the quark scalar-density ratio, $C_B(\bar{\sigma})$, in the linear form~\cite{QMC,Tsushima:1997cu}
\bge
	C_{B}(\bar{\sigma}) = 1 - a_{B}\times(g_{\sigma N}\bar{\sigma}), 
	\label{eq:Csigma}
\ene
where $g_{\sigma N}$ is the $\sigma$-$N$ coupling constant at zero density and $a_B$ is a parameter.  
Using this parameterization, 
the $\sigma$-$B$ coupling constant in Eq.~(\ref{eq:Lagrangian-interaction}) is given by~\cite{QMC,Tsushima:1997cu}
\bge
	g_{\sigma B}(\bar{\sigma}) = g_{\sigma B}b_{B}\left[1-\frac{a_{B}}{2}(g_{\sigma N}\bar{\sigma})\right],
	\label{sigma-coupling-const}
\ene
where we introduce a parameter $b_{B}$ to use Eq.~(\ref{sigma-coupling-const}) in both the QMC and CQMC models. 
The values of $a_B$ and $b_B$ are listed in 
Table~\ref{tab:parametrizationQMC}.  
If we set $a_{B}=0$ and $b_{B}=1$, $g_{\sigma B}(\bar{\sigma})$ is identical to the $\sigma$-$B$ coupling constant 
in QHD. 
\begin{table}
\caption{\label{tab:parametrizationQMC}
Values of $a_B$ and $b_B$ for various baryons in the QMC or CQMC model. }
\begin{ruledtabular}
\begin{tabular}{lcccc}
\ & QMC & \ & CQMC & \ \\
\ & $a_{B}$~(fm) & $b_{B}$ & $a_{B}$~(fm) & $b_{B}$ \\
\hline
$N$       & 0.179 & 1.00 & 0.118 & 1.04 \\
$\Lambda$ & 0.172 & 1.00 & 0.122 & 1.09 \\
$\Sigma$  & 0.177 & 1.00 & 0.184 & 1.02 \\
$\Xi$     & 0.166 & 1.00 & 0.181 & 1.15 \\
\end{tabular}
\end{ruledtabular}
\end{table}

In the present calculation, we add a non-linear (NL) term of the $\bar{\sigma}$,
\bge
	U(\bar{\sigma}) = \frac{g_{2}}{3}\bar{\sigma}^{3} + \frac{g_{3}}{4}\bar{\sigma}^{4}  , 
	\label{eq:scalar-self-interaction}
\ene
to {\em only} the QHD Lagrangian, because the EOS given by the {\em naive} QHD is too hard~\cite{Serot:1984ey}.  
We call this QHD+NL. 

To sum up all orders of the tadpole and exchange diagrams in the baryon Green's function, $G_B$, 
we use the Dyson's equation 
\bge
	G_{B}(k) = G_{B}^{0}(k) + G_{B}^{0}(k)\Sigma_{B}(k)G_{B}(k) , 
	\label{eq:Dyson-equation}
\ene
where $\Sigma_{B}$ is the baryon self-energy and $G_{B}^{0}$ is the Green's function for the free baryon. 
In nuclear or neutron matter, 
the baryon self-energy is generally written as
\bge
	\Sigma_{B}(k) = \Sigma_{B}^{s}(k) - \gamma_{0}\Sigma_{B}^{0}(k) 
	+ (\vec{\gamma}\cdot\hat{k})\Sigma_{B}^{v}(k),
	\label{eq:baryon-self-engy}
\ene
with $\hat{k}$ the unit vector along the momentum $\vec{k}$ and $\Sigma_{B}^{s(0)[v]}$ 
the scalar part (time component) [space component] of the self-energy. 
Furthermore, the effective baryon mass, momentum and energy in matter are respectively defined by~\cite{Bouyssy:1987sh}
\bg
	M_{B}^{\ast}(k) &=& M_{B} + \Sigma_{B}^{s}(k),\\
	k_{B}^{\ast\mu} &=& (k_{B}^{\ast0},\vec{k}_{B}^{\ast}) 
	= (k^{0}+\Sigma_{B}^{0}(k),\vec{k}+\hat{k}\Sigma_{B}^{v}(k)) ,\\
	E_{B}^{\ast}(k) &=& \left[\vec{k}_{B}^{\ast2}+M_{B}^{\ast2}(k)\right]^{1/2}.
	\label{eq:auxiliary-quantity}
\en
\begin{table}
\caption{\label{tab:BSE} Functions $A_{i}$, $B_{i}$, $C_{i}$, and $D_{i}$.  
The index $i$ is specified in the left column, where $V (T)$ stands for the vector (tensor) coupling at each
meson-$BB^\prime$ vertex.  The last row is for the (pseudovector) pion contribution. 
}
\begin{ruledtabular}
\begin{tabular}{lcccc}
$i$ & $A_{i}$ & $B_{i}$ & $C_{i}$ & $D_{i}$ \\
\hline
$\sigma$ & 
$g_{\sigma B}^{2}(\bar{\sigma})\Theta_{\sigma}$ & 
$g_{\sigma B}^{2}(\bar{\sigma})\Theta_{\sigma}$ & 
$-2g_{\sigma B}^{2}(\bar{\sigma})\Phi_{\sigma}$ & 
$-$ \\
$\omega_{VV}$ & 
$2g_{\omega B}^{2}\Theta_{\omega}$ & 
$-4g_{\omega B}^{2}\Theta_{\omega}$ & 
$-4g_{\omega B}^{2}\Phi_{\omega}$ & 
$-$ \\
$\omega_{TT}$ & 
$-\left(\frac{f_{\omega B}}{2\mathcal{M}}\right)^{2}m_{\omega}^{2}\Theta_{\omega}$ & 
$-3\left(\frac{f_{\omega B}}{2\mathcal{M}}\right)^{2}m_{\omega}^{2}\Theta_{\omega}$ & 
$4\left(\frac{f_{\omega B}}{2\mathcal{M}}\right)^{2}\Psi_{\omega}$ & 
$-$ \\
$\omega_{VT}$ & 
$-$ & 
$-$ & 
$-$ & 
$12\frac{f_{\omega B}g_{\omega B}}{2\mathcal{M}}\Gamma_{\omega}$ \\
$\rho_{VV}$ & 
$2g_{\rho B}^{2}\Theta_{\rho}$ & 
$-4g_{\rho B}^{2}\Theta_{\rho}$ & 
$-4g_{\rho B}^{2}\Phi_{\rho}$ &
$-$ \\
$\rho_{TT}$ & 
$-\left(\frac{f_{\rho B}}{2\mathcal{M}}\right)^{2}m_{\rho}^{2}\Theta_{\rho}$ & 
$-3\left(\frac{f_{\rho B}}{2\mathcal{M}}\right)^{2}m_{\rho}^{2}\Theta_{\rho}$ & 
$4\left(\frac{f_{\rho B}}{2\mathcal{M}}\right)^{2}\Psi_{\rho}$ & 
$-$ \\
$\rho_{VT}$ & 
$-$ & 
$-$ & 
$-$ & 
$12\frac{f_{\rho B}g_{\rho B}}{2\mathcal{M}}\Gamma_{\rho}$ \\
$\pi_{pv}$ &
$-f_{\pi B}^{2}\Theta_{\pi}$ & 
$-f_{\pi B}^{2}\Theta_{\pi}$ & 
$2\left(\frac{f_{\pi B}}{m_{\pi}}\right)^{2}\Pi_{\pi}$ & 
$-$ \\
\end{tabular}
\end{ruledtabular}
\end{table}
The baryon self-energies in Eq.~(\ref{eq:baryon-self-engy}) are then calculated by~\cite{Bouyssy:1987sh}
\bg
	\Sigma_{B}^{s}(k) &=& -g_{\sigma B}(\bar{\sigma})\bar{\sigma}
	+ \sum_{B^{\prime}, i} \frac{I_{BB^{\prime}}^{i}}{(4\pi)^{2}k}\int_{0}^{k_{F_{B^{\prime}}}} dq \, q
	\left[\frac{M_{B^{\prime}}^{\ast}(q)}{E_{B^{\prime}}^{\ast}(q)}B_{i}(k,q)
	+ 
	\frac{q_{B^{\prime}}^{\ast}}{2E_{B^{\prime}}^{\ast}(q)}D_{i}(q,k)\right],
	\label{eq:BSE-scalar} \\
	\Sigma_{B}^{0}(k) &=& -g_{\omega B}\bar{\omega}-g_{\rho B}({\vec I}_{B})_3\bar{\rho}
	-\sum_{B^{\prime}, i} \frac{I_{BB^{\prime}}^{i}}{(4\pi)^{2}k} \int_{0}^{k_{F_{B^{\prime}}}}dq \, q
	A_{i}(k,q),
	\label{eq:BSE-time} \\
	\Sigma_{B}^{v}(k) &=& \sum_{B^{\prime}, i} \frac{I_{BB^{\prime}}^{i}}{(4\pi)^{2}k} 
	\int_{0}^{k_{F_{B^{\prime}}}}dq \, q
	\left[\frac{q_{B^{\prime}}^{\ast}}{E_{B^{\prime}}^{\ast}(q)}C_{i}(k,q)
	+ 
	\frac{M_{B^{\prime}}^{\ast}(q)}{2E_{B^{\prime}}^{\ast}(q)}D_{i}(k,q)\right],
	\label{eq:BSE-vector}
\en
where $\bar{\omega}$ and $\bar{\rho}$ are respectively the mean-field values of ${\omega}$ and ${\rho}$ mesons 
in matter, $k_{F_B}$ is the Fermi momentum for baryon $B$ 
and the factor, $(I_{BB^{\prime}}^{i})^{1/2}$, is the isospin weight at the meson-$BB^\prime$ vertex 
in the Fock diagram. 
The index $i$ in the sum and the functions $A_{i}$, $B_{i}$, $C_{i}$ and $D_{i}$ 
in Eqs.(\ref{eq:BSE-scalar})-(\ref{eq:BSE-vector}) are explicitly given in Table~\ref{tab:BSE}, 
in which the following functions are used~\cite{Bouyssy:1987sh}:
\bg
	\Theta_{i}(k,q) &=& \ln\left[\frac{m_{i}^{2}+(k+q)^{2}}{m_{i}^{2}+(k-q)^{2}}\right], \\
	\Phi_{i}(k,q) &=& \frac{1}{4kq}\left(k^{2}+q^{2}+m_{i}^{2}\right)\Theta_{i}(k,q)-1,\\
	\Psi_{i}(k,q) &=& \left[\left(k^{2}+q^{2}-m_{i}^{2}/2\right)\Phi_{i}(k,q)-kq\Theta_{i}(k,q)\right], \\
	\Pi_{i}(k,q) &=& \left[\left(k^{2}+q^{2}\right)\Phi_{i}(k,q)-kq\Theta_{i}(k,q)\right], \\
	\Gamma_{i}(k,q) &=& \left[k\Theta_{i}(k,q)-2q\Phi_{i}(k,q)\right].
\en

The mean-field values of $\bar{\omega}$ and $\bar{\rho}$ are respectively given by the usual forms 
\bge
	\bar{\omega} = \sum_{B}\frac{g_{\omega B}}{m_{\omega}^{2}}\rho_{B}, \ \ \ \ \ 
	\bar{\rho} = \sum_{B}\frac{g_{\omega B}}{m_{\omega}^{2}}(I_{B})_3\rho_{B},
	\label{eq:rho-field}
\ene
where the density of baryon $B$ is $\rho_{B} = (\frac{2J_{B}+1}{6\pi^{2}}){k^{3}_{F_{B}}}$ with $J_B$ the spin of $B$. 

On the other hand, combining with Eqs.(\ref{eq:BSE-scalar})-(\ref{eq:BSE-vector}), the value of $\bar{\sigma}$ is self-consistently calculated by~\cite{RHF-Huber}
\bge
	\bar{\sigma} = \sum_{B}\frac{g_{\sigma B}}{m_{\sigma}^{2}}b_{B}C_{B}(\bar{\sigma})\rho_{B}^s
	- \frac{1}{m_{\sigma}^{2}}(g_{2}\bar{\sigma}^{2}+g_{3}\bar{\sigma}^{3}) , 
	\label{eq:sigma-field}  
\ene
where the scalar density of baryon $B$ reads 
\bge
\rho_{B}^s = \frac{2J_{B}+1}{2\pi^{2}}\int_{0}^{k_{F_{B}}}dk~k^{2}\frac{M_{B}^{\ast}(k)}{E_{B}^{\ast}(k)} . 
	\label{eq:baryon-scalar-density} 
\ene
In Eq.(\ref{eq:sigma-field}), $g_2$ and $g_3$ vanish in the QMC or CQMC model, while, 
in QHD+NL, $b_BC_{B}(\bar{\sigma})$ is taken to be unity. 

The total energy density (pressure) for matter can be divided into the baryon and lepton contributions, 
namely $\epsilon=\epsilon_{B}+\epsilon_{\ell}$ ($P=P_{B}+P_{\ell}$), 
where the baryon energy density, $\epsilon_{B}$, and pressure, $P_{B}$, are expressed by
\bge
	\epsilon_{B} = \sum_{B}\frac{2J_{B}+1}{(2\pi)^{3}}\int_{0}^{k_{F_{B}}}d\vec{k}~\left[
	T_{B}(k)+\frac{1}{2}V_{B}(k)\right]
	-\frac{\bar{\sigma}^{3}}{2}\left(\frac{g_{2}}{3}+\frac{g_{3}}{2}\bar{\sigma}\right),
	\label{eq:baryon-engy-density}
\ene
with
\bge
	T_{B}(k) = \frac{M_{B}M_{B}^{\ast}(k)+kk_{B}^{\ast}}{E_{B}^{\ast}(k)} , \ \ \ \ \ 
	V_{B}(k) = \frac{M_{B}^{\ast}(k)\Sigma_{B}^{s}(k)+k_{B}^{\ast}\Sigma_{B}^{v}(k)}{E_{B}^{\ast}(k)}-\Sigma_{B}^{0}(k)
	\label{eq:baryon-engy-density-kinetic-potential} , 
\ene
and 
\bge
	P_{B} = n_{B}^{2}\frac{\partial}{\partial n_{B}}\left(\frac{\epsilon_{B}}{n_{B}}\right)
	\label{eq:baryon-pressure} . 
\ene
In Eq.(\ref{eq:baryon-pressure}), the total baryon density, $n_{B}$, is given by $n_{B} = \sum_{B}\rho_{B}$.

The numerical result for the properties of symmetric nuclear matter is presented in Table~\ref{tab:cc}.  
In the present calculation, 
the $\sigma$-$N$ and $\omega$-$N$ coupling constants are determined so as 
to reproduce the saturation energy ($-15.7$ MeV) at normal nuclear density ($n_B^0=0.15$ fm$^{-3}$). 
\begin{table}
\caption{\label{tab:cc}Coupling constants and calculated properties of symmetric nuclear matter at $n_B^0$.  
In the left column, H (HF) stands for the Hartree (Hartree-Fock) calculation. 
The incompressibility, $K_{v}$, 
and the symmetry energy, $a_{4}$, are in MeV.
The coupling constants, $g_{2}$ (in fm$^{-1}$) and $g_{3}$, in QHD+NL 
are chosen so as to adjust $K_{v}$ and $M_{N}^{\ast}$ to the values given by the QMC model. 
} 
\begin{ruledtabular}
\begin{tabular}{lcccccccc}
Model & $g_{\sigma N}^{2}/4\pi$ & $g_{\omega N}^{2}/4\pi$ & $g_{\rho N}^{2}/4\pi$ & $g_{2}$ & $g_{3}$ 
& $K_{v}$ & $M_{N}^{\ast}/M_{N}$ & $a_{4}$ \\
\hline
QHD+NL(H)   & 5.34 & 5.41 & 1.88 & 11.07 & 95.59 & 280 & 0.80 & 36.9 \\
QMC(H)      & 5.45 & 5.41 & 1.88 & $-$   & $-$   & 280 & 0.80 & 36.9 \\
CQMC(H)     & 5.75 & 7.11 & 1.82 & $-$   & $-$   & 302 & 0.76 & 36.9 \\
\hline
QHD+NL(HF)  & 3.63 & 6.66 & 0.48
& 16.27 & 48.52 & 280 & 0.73 & 38.6 \\
QMC(HF)     & 3.50 & 6.41 & 0.48 & $-$ & $-$ & 280 & 0.73 & 36.9 \\
CQMC(HF)    & 3.59 & 7.34 & 0.48 & $-$ & $-$ & 300 & 0.70 & 37.4 \\
\end{tabular}
\end{ruledtabular}
\end{table}

In the Hartree calculation, the $\rho$-$N$ coupling constant, $g_{\rho N}$, 
is determined so as to fit the symmetry energy, $a_{4}=36.9$ MeV~\cite{Shen:2002qg,Ishizuka:2008gr,Sugahara:1993wz}, and 
the meson-hyperon coupling constants, $g_{\sigma Y}$, $g_{\omega Y}$ and $g_{\rho Y}$, are taken to be the values 
derived from the quark model~\cite{Pal:1999sq,Shen:2002qg}.  

In the Hartree-Fock calculation, we use the more reliable set of the coupling constants 
(except $g_{\sigma B}$ and $g_{\omega N}$) 
determined in the Nijmegen extended-soft-core (ESC) model (see Table~VII in Ref.~\cite{Rijken:2010zzb}). 
Furthermore, from the recent analyses of hypernuclei and hyperon production reactions, it is inferred that 
the $\Lambda \, (\Sigma) \, [\Xi]$ feels the potential, 
$U_{\Lambda \, (\Sigma) \, [\Xi]} \simeq -30 \, (+30) \, [-15]$ MeV, 
in nuclear matter~\cite{Ishizuka:2008gr,Tsushima:1997cu}.  Therefore, we adjust the coupling constants, 
$g_{\sigma \Lambda}$, $g_{\sigma \Sigma}$ and $g_{\sigma \Xi}$, 
so as to reproduce the suggested potential values at $n_B^0$. In the case of QHD+NL (QMC) [CQMC], 
we take $g_{\sigma \Lambda}/\sqrt{4\pi}=2.88 \, (2.78) \, [2.44]$, 
$g_{\sigma \Sigma}/\sqrt{4\pi}=2.05 \, (1.97) \, [1.94]$ 
and $g_{\sigma \Xi}/\sqrt{4\pi}=2.06 \, (1.98) \, [1.70]$. 

\begin{table}
\caption{\label{tab:nucleaon-self-engy}
Nucleon self-energies, $\Sigma_{N}^{s,0,v}$, at $n_B^0$ in symmetric nuclear matter.  
The values of the self-energy are in MeV.}
\begin{ruledtabular}
\begin{tabular}{llrrr}
\ & \ & $\Sigma_{N}^{s}$ & $\Sigma_{N}^{0}$ & $\Sigma_{N}^{v}$ \\
\hline
Hartree      & \        & -186 & -128 & 0 \\
\hline
Hartree-Fock & \        & \    & \    & \  \\
Direct       & \        & -131 & -151 &  0 \\
Exchange     & $\sigma$ &   13 &  -14 &  0 \\
\            & $\omega$ &  -63 &  -32 & -3 \\
\            & $\pi$    &   -4 &    4 & -4 \\
\            & $\rho$   &  -69 &   13 & 14 \\
total        & \        & -255 & -180 &  6 \\
\end{tabular}
\end{ruledtabular}
\end{table}
In Table~\ref{tab:nucleaon-self-engy}, the contents of the nucleon self-energies in the QMC model are presented. 
The Fock term contributes to the self-energy significantly.  
In symmetric nuclear matter, the effect of the tensor coupling of $\omega$ meson is small, while that of 
the $\rho$ meson is considerable; for example, the $\rho$-meson contributions of the vector ($VV$), tensor ($TT$) 
and vector-tensor ($VT$) mixing  
to the scalar part, $\Sigma_N^s$, are respectively $-14$ MeV, $-58$ MeV and $4$ MeV. 
The space component, $\Sigma_{N}^{v}$, at $n_B^0$ is relatively small. 

\begin{figure}
\includegraphics[width=250pt,keepaspectratio,clip,angle=270]{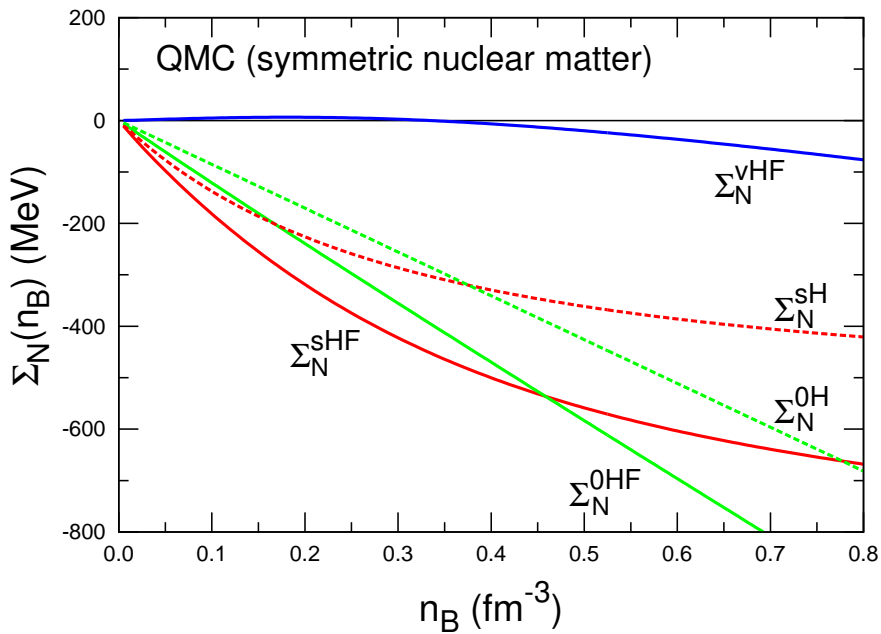}%
\caption{\label{fig:nucleaon-self-engy} 
Nucleon self-energies, $\Sigma_N^{s,0,v}$, in symmetric nuclear matter.  
}
\end{figure}
In Fig.~\ref{fig:nucleaon-self-engy}, 
we show the nucleon self-energies calculated by the QMC model as functions of the total baryon density, $n_B$.   
The scalar and time components, $\Sigma_{N}^{s,0}$, in the Hartree-Fock calculation are much deeper than 
those in the Hartree calculation.  
It is noticeable that the space component also becomes deep at high densities 
and thus it is not negligible in dense matter. 

\begin{figure}
\includegraphics[width=250pt,keepaspectratio,clip,angle=270]{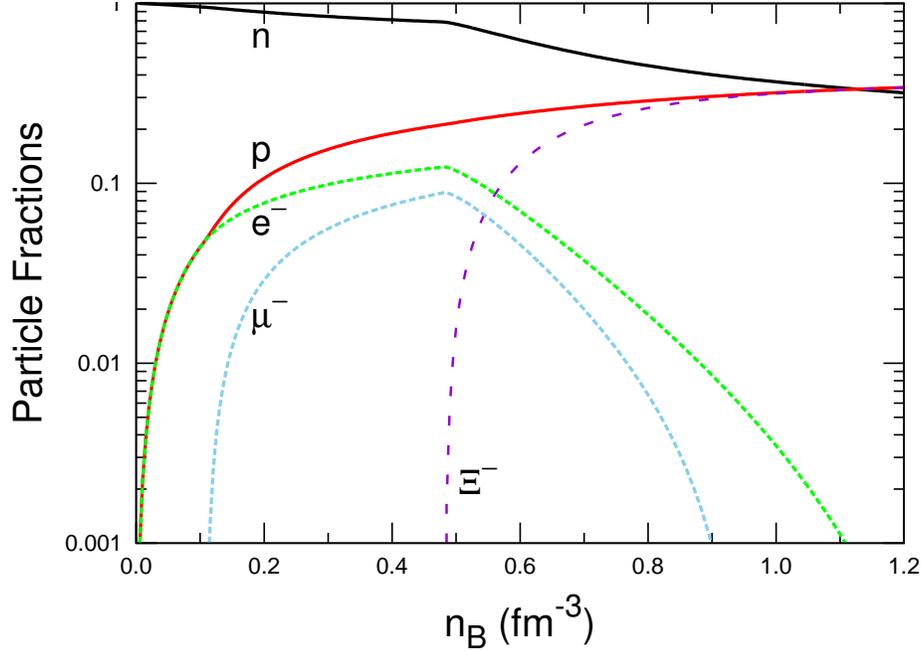}%
\caption{\label{fig:Composition} 
Particle fractions in neutron matter.}
\end{figure}
\begin{table}
\caption{\label{tab:NSproperty}
Neutron-star radius, $R_{max}$ (in km), the central density, $n_{c}$ (in fm$^{-3}$), and the 
ratio of the maximum neutron-star mass to the solar mass, $M_{max}/M_{\odot}$. 
The Hartree and the Hartree-Fock calculation with (without) hyperons are denoted by npY (np). }
\begin{ruledtabular}
\begin{tabular}{l|ccc|ccc}
\ & \ & np & \ & \ & npY & \ \\
\ & $R_{max}$ & $n_{c}$ & $M_{max}/M_{\odot} $ & $R_{max}$ & $n_{c}$ & $M_{max}/M_{\odot} $ \\
\hline
QHD+NL(H)  & 11.3 & 1.04 & 2.00 & 12.5 & 0.86 & 1.56 \\
QMC(H)     & 11.5 & 1.01 & 2.05 & 12.5 & 0.86 & 1.60 \\
CQMC(H)    & 11.8 & 0.92 & 2.20 & 12.5 & 0.88 & 1.66 \\
\hline
QHD+NL(HF) & 11.7 & 0.95 & 2.15 & 11.9 & 0.95 & 1.92 \\
QMC(HF)    & 11.5 & 0.97 & 2.11 & 12.0 & 0.92 & 1.95 \\
CQMC(HF)   & 11.9 & 0.90 & 2.23 & 12.3 & 0.87 & 2.02 \\
\end{tabular}
\end{ruledtabular}
\end{table}
\begin{figure}
\includegraphics[width=250pt,keepaspectratio,clip,angle=270]{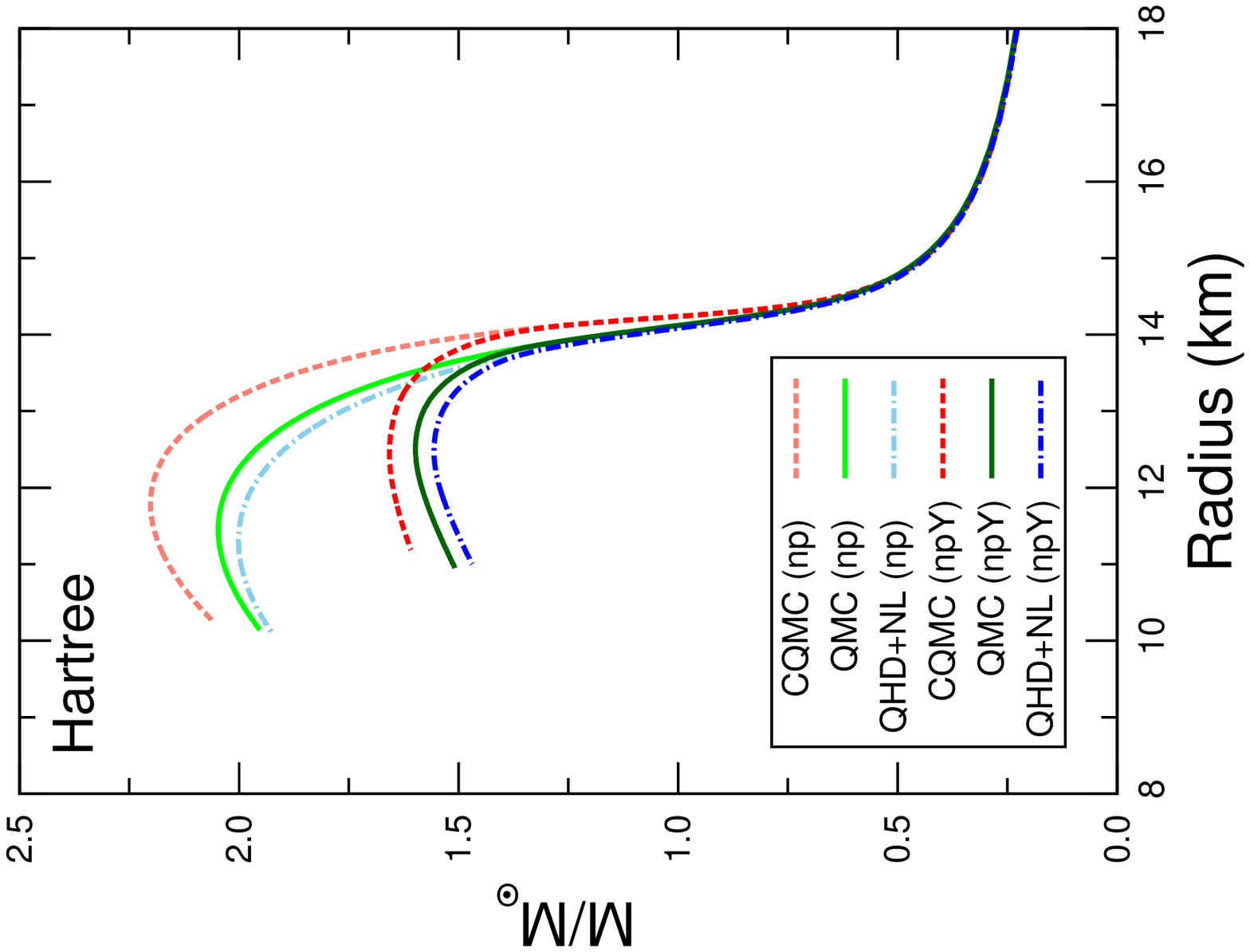}%
\ 
\includegraphics[width=250pt,keepaspectratio,clip,angle=270]{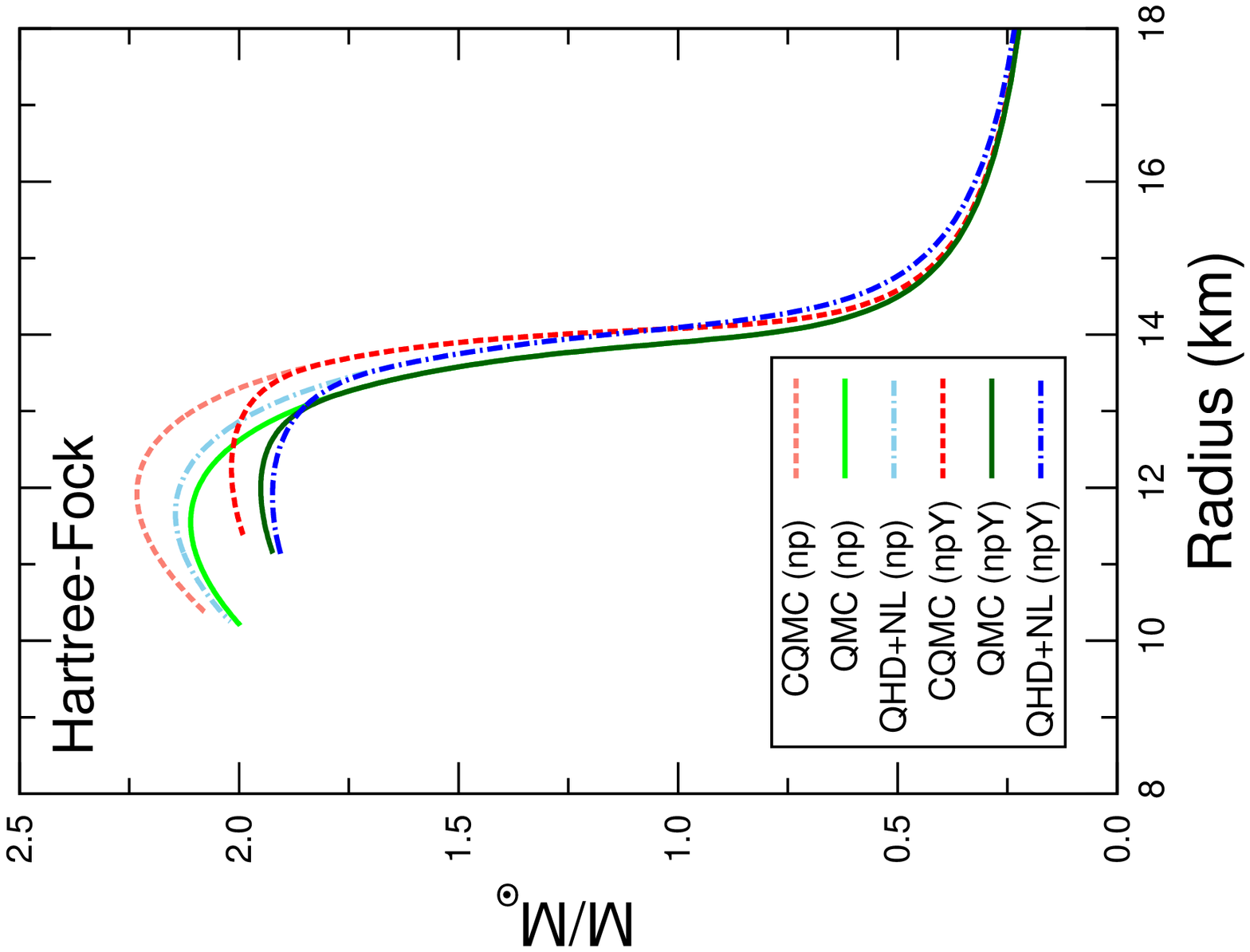}%
\caption{\label{fig:TOV} 
Neutron-star mass as a function of the radius. 
The left (right) panel is for the Hartree (Hartree-Fock) calculation.   In solving the TOV equation, 
we use the EOS of BBP~\cite{Baym:1971ax} and BPS~\cite{Baym:1971pw} at very low densities.  
}
\end{figure}
In a neutron star, the charge neutrality and the $\beta$ equilibrium under weak processes are realized. 
Under these conditions, we calculate the EOS for neutron matter and solve the 
Tolman-Oppenheimer-Volkoff (TOV) equation. 
In Fig.~\ref{fig:Composition}, we show the QMC result of particle fractions in relativistic Hartree-Fock approximation.  
With respect to hyperons, only 
the $\Xi^{-}$ appears around $0.49$ fm$^{-3}$ and 
the other hyperons are {\em not} produced at densities below 1.2 fm$^{-3}$. 
We note that, in the case of QHD+NL, the $\Xi^-$ first appears around  $0.43$ fm$^{-3}$, 
and that the $\Lambda$ and $\Xi^0$ are produced at densities beyond $0.69$ fm$^{-3}$.  
This tendency is similar to the result of Ref.~\cite{RikovskaStone:2006ta}. 
It may be interesting to compare the present result with the Hartree result given in Ref.~\cite{Carroll:2008sv}. 

We summarize the properties of neutron star in Table~\ref{tab:NSproperty}, 
and show the mass as a function of the neutron-star radius in Fig.~\ref{fig:TOV}. 
As known well, the inclusion of hyperons generally reduces the mass of a neutron star. 
However, because the Fock contribution makes the EOS hard, the maximum mass in the present calculation can reach   
the recently observed value, $1.97\pm0.04 M_{\odot}$.  
If we ignore the tensor coupling in the Fock term, the difference between the maximum masses 
in the Hartree and the Hartree-Fock calculations is not large. 
Therefore, the tensor coupling (especially, in the high density region) is very vital to obtain the large neutron-star mass. 
The variation of the quark substructure of baryon in matter also enhances the mass, 
because the QMC model includes the (repulsive) many-body interaction given through 
the scalar polarizability in the coupling constant, $g_{\sigma B}(\bar{\sigma})$~\cite{Saito:2010zw,Guichon:2006er}. 
Furthermore, the quark-quark hyperfine interaction due to the gluon and pion exchanges increases 
the mass of a neutron star.  We should note that  
the inclusion of the meson-baryon form factor at the interaction vertex 
in the Fock term does {\em not} change the present result much. 

In summary, we have studied the effects of the Fock term, 
the tensor couplings of vector mesons and the baryon structure variation on the properties of a neutron star. 
The present calculation can produce the maximum neutron-star mass of $\sim 2.0 M_\odot$, 
which is consistent with the recently observed mass, $1.97\pm0.04 M_{\odot}$. 
The Fock contribution is very important and, particularly, 
the inclusion of tensor coupling is necessary to obtain such large mass. 
The in-medium variation of baryon structure also makes the EOS hard and thus it enhances the mass of a neutron star. 
In the future, it may be desirable to consider the degrees of freedom of 
$\Delta$ isobar~\cite{Glendenning:1997wn,RHF-Huber}, 
the $K$-meson exchange and the baryon mixing in the Fock diagram. 
It is also interesting to study the possibility of meson condensates~\cite{Glendenning:1997wn,Schaffner:1995th,Meson-condensation} 
and/or quark matter~\cite{Glendenning:1997wn}. 

\vspace{1cm}
The authors would like to thank K. Nakazato, C. Ishizuka and H. Suzuki 
for fruitful discussions on the equation of state for a neutron star.

%
%

%

\newpage 

\end{document}